\documentclass[aps,showpacs,twocolumn]{revtex4}
\usepackage{amsmath, amsthm,times}
\usepackage{epsfig}
\usepackage{amssymb}
\usepackage{graphicx}

\begin{document}

\sloppy

\title{Strengthened Bell Inequalities for Entanglement Verification}

\author{Pavel Lougovski$^{1}$ and S.J. van Enk$^{1,2}$}

\address{
$^1$Department of Physics and Oregon Center for Optics, University of Oregon\\
 Eugene, OR 97403\\
$^2$Institute for Quantum Information, California Institute of Technology, Pasadena, CA 91125}
\begin{abstract}
Bell inequalities were meant to test quantum mechanics vs local hidden variable models, but can also be used to verify entanglement. For entanglement verification purposes one assumes the validity of quantum mechanics as well as quantum descriptions of one's measurements. With the help of these assumptions it is possible to derive a {\em strengthened} Bell inequality whose violation implies entanglement.
We generalize known examples of such inequalities by relating the
expectation value of the Bell operator to a particular quantitative measure of entanglement, namely the negativity. Moreover, we obtain statistics illustrating the fact that violating a given (strengthened or not) Bell inequality is a much more rare feat for a quantum state of two qubits than it is to be entangled. 
\end{abstract}

\pacs{03.67.Mn, 03.65.Ud}

\maketitle

Bell introduced his famous inequality as a way of testing predictions of quantum mechanics versus those of all possible local hidden variable (LHV) theories \cite{bell}. But the inequality (or its experimentally useful generalization, the CHSH inequality \cite{chsh}) can also be used to verify the presence of entanglement under the specific assumption that quantum mechanics is correct, if the inequality is violated. In fact, since the Bell-CHSH inequalities are derived from classical probability theory, without depending at all on quantum mechanics, such tests are safe in the sense of avoiding pitfalls resulting from assuming too much about the quantum state generated \cite{elk}.

Let us consider a scenario where (a multitude of copies of) a quantum state $\rho$ is shared by two parties A and B. Moreover, each party has a choice of two local measurements with dichotomic outcomes $\pm 1$, say, $A_{1},A_{2}$ for party A and $B_{1},B_{2}$ for party B.  Then one can define the Bell-CHSH~\cite{chsh} operator,
\begin{equation}
\label{CHSHoper}
\mathcal{B} = A_{1}\otimes(B_{1} + B_{2}) + A_{2}\otimes(B_{1} - B_{2}).
\end{equation}
 In order to calculate the expectation value of $\mathcal{B}$ for a quantum state $\rho$ the standard quantum-mechanical rule should be applied, i.e. 
 \begin{equation}
 \langle\mathcal{B}\rangle_{QM}=\rm{Tr}(\mathcal{B}\rho).
\end{equation}
 On the other hand, if we assume that a LHV model is correct, then expectation values should be calculated differently [and the tensor product in (\ref{CHSHoper}) should be replaced by an ordinary product], namely by 
 \begin{equation}
\langle\mathcal{B}\rangle_{LHV}=\int_{\Omega}d\omega p(\omega)\mathcal{B}(\omega),
\end{equation}
 where $\omega$ is a set of "hidden" variables and $p(\omega)$ is the probability density for those variables.
The CHSH inequality reads
\begin{equation}\label{BLHV}
|\langle\mathcal{B}\rangle_{LHV}| \le 2.
\end{equation}
This inequality holds true for all LHV models, but can be violated by some quantum-mechanical states for certain choices of operators $A_{1,2}, B_{1,2}$. That is, there are states for which $|\langle\mathcal{B}\rangle_{QM}| > 2$.
Such quantum states must be entangled, as all separable states satisfy the inequality. Hence, the violation of the CHSH inequality provides a sufficient criterion for entanglement.  

The largest possible violation of the CHSH inequality allowed by quantum mechanics is given by Cirel'son's bound~\cite{Cirel'son}:
\begin{equation}\label{BQM}
|\langle\mathcal{B}\rangle_{QM}| \le 2\sqrt2.
\end{equation}
 On the other hand, it was recently shown by Uffink and Seevinck~\cite{UffinkSeevinck} that, in the special case that the measurements $A_{1,2}$ and $B_{1,2}$ correspond to spin measurements  in locally orthogonal spatial directions \footnote{No relation is assumed between the directions in location $A$ and those in $B$.} (on spin-1/2 systems, which we will refer to as qubits),  a significantly stronger inequality can be found for separable states. From that inequality one can derive a ``strengthened Bell inequality'' 
 \begin{equation}\label{RUS}
|\langle\mathcal{B}\rangle_{QM,sep., \perp}| \le \sqrt 2,
\end{equation}
 (derived explicitly first, as far as we know, by Roy in Ref.~\cite{roy}; the inequality is also implicit in, e.g., Ref.~\cite{gurk}). We will refer to this inequality as the Roy-Uffink-Seevinck (RUS) bound. The subscript $\perp$ reminds us that it holds only for spin measurements in orthogonal directions, and the subscript $sep$ reminds us the inequality refers to separable states. The RUS bound demonstrates the known fact that one certainly does not have to violate the CHSH inequality (\ref{BLHV}) in order to conclude one has entanglement.
In other words, all data from an experiment 
measuring the Bell correlations in which the RUS bound is violated, but the CHSH inequality is not, can be 
reproduced by a LHV model, although the underlying quantum state {\em must} be entangled, if orthogonal spin directions were measured.

We now generalize the RUS and Cirel'son's bounds.
We consider, after Uffink and Seevinck~\cite{UffinkSeevinck}, orthogonal spin measurements $\{A_{1},A_{2}\}$ and $\{B_{1},B_{2}\}$ on a spin-1/2 system (qubit) for each party. For this choice of local measurements the Bell-CHSH operator $\mathcal{B}$ in Eq.(\ref{CHSHoper}) has only two non-zero eigenvalues $\pm 2\sqrt2$, with the corresponding eigenvectors being two orthogonal maximally-entangled two-qubit states. Let us now maximize the expectation value of $\mathcal{B}$ over all possible pairs of orthogonal spin measurement directions,
\begin{equation}
\langle\mathcal{B}\rangle_{\rm{max}} =\rm{max}\{\rm{Tr}(U\mathcal{B}U^{-1}\rho), U=U_{A}\otimes U_{B}\}.\end{equation}
Calculating the trace in the basis where $\mathcal{B}$ is diagonal we immediately obtain
\begin{equation}
\langle\mathcal{B}\rangle_{\rm{max}} = \rm{max}(2\sqrt2(\langle\psi|U^{-1}\rho U|\psi\rangle - \langle\phi|U^{-1}\rho U|\phi\rangle)),
\end{equation}
where $|\psi\rangle$ and $|\phi\rangle$ are maximally-entangled states, the eigenstates of $\mathcal{B}$ with eigenvalues $\pm2\sqrt{2}$. Now we recall that for an arbitrary two-qubit quantum state $\rho$ we can define a fidelity $F$~\cite{BDSW}
\begin{equation}
F(\rho) = \rm{max}\langle\Psi|U_{A}\otimes U_{B}\rho (U_{A}\otimes U_{B})^{-1}|\Psi\rangle,
\end{equation}
where $|\Psi\rangle$ is a maximally-entangled state and $U_{A},U_{B}$ are local unitaries. Combining these two definitions shows  that the absolute value of $\langle\mathcal{B}\rangle_{\rm{max}}$ is bounded from above by 
\begin{equation}
|\langle\mathcal{B}\rangle_{\rm{max}}|\le2\sqrt 2 F(\rho).
\end{equation}
 Furthermore, for any mixed two-qubit state $\rho$ we have the inequality  (proven in Ref.~\cite{FidelityNeg})
 \begin{equation}
 F(\rho)\le(1+N(\rho))/2,
 \end{equation}
 where $N(\rho)$ is the negativity~\cite{NegativityOriginal,VidalWerner}, defined as
\begin{equation}
N(\rho)=2\sum_k\max(0,-\lambda_k),
\end{equation} 
in terms of the eigenvalues $\lambda_k$ of the partial transpose of $\rho$. We thus arrive at the announced generalization of the RUS inequality for two-qubit states:
\begin{equation}
\label{newbound}
|\langle\mathcal{B}\rangle_{QM,\perp}|\le\sqrt 2 (1+N(\rho)).
\end{equation}
It is straightforward to see that the last inequality, Eq.~(\ref{newbound}), contains both the RUS and Cirel'son's bounds. Namely, when $N(\rho)=0$, $\rho$ is a separable state (this is true for our two-qubit states, although not in general) and we recover the RUS bound. On the other hand, for maximally-entangled states $N(\rho)=1$ and  Cirel'son's bound is recovered.
Moreover, Eq.(\ref{newbound}) provides a lower bound on the degree of entanglement in terms of negativity, if one violates the RUS bound. The bound complements the relations found in \cite{wolf} between Bell inequalities and the concurrence and purity of mixed states.

The next question we address is this: if one measures spin in two fixed orthogonal spatial directions, then what is the probability that a randomly picked two-qubit state violates the original CHSH inequality? Or what is the probability it violates the RUS bound? How do these two probabilities compare to the probability of the state being entangled? 

In order to answer these questions, we have to specify how to ``randomly'' pick a two-qubit state. 
In fact, this boils down to choosing a measure on the space of all two-qubit density matrices. This choice is rather arbitrary. 
The only (rather weak) property we wish the random distribution of states to have, is that a nonnegligble fraction is entangled, and that a nonnegligible fraction is separable.
We pick then, for no other reason in particular,
a parametrization  based on Ref.~\cite{NegativityOriginal} (see also \cite{TilmaSudarshan}).  
A random sample of $3\times 10^7$ states showed a fraction $(36.437 \pm 0.010)\%$ to be entangled {\em a priori} (as compared to $(36.8\pm 0.2)\%$ obtained in \cite{NegativityOriginal} for the same measure), thus fulfilling our one weak condition.

The following statistics apply to a {\em fixed} set of {\em four} correlation measurements involving spin measurements in two {\em fixed} orthogonal spatial directions on each of the two qubits. 
We note that in this case we can construct four Bell-CHSH operators from the four measured correlations:
\begin{eqnarray}
\mathcal{B}_1&:=& A_{1}\otimes(B_{1} + B_{2}) +A_{2}\otimes(B_{1} - B_{2}), \nonumber\\
\mathcal{B}_2&=& A_{1}\otimes(B_{1} + B_{2}) -A_{2}\otimes(B_{1} - B_{2}), \nonumber\\
\mathcal{B}_3&=& A_{1}\otimes(B_{1} - B_{2}) + A_{2}\otimes(B_{1} + B_{2}), \nonumber\\
\mathcal{B}_4&=& A_{1}\otimes(-B_{1} + B_{2}) + A_{2}\otimes(B_{1} + B_{2}).\label{AllBell}
\end{eqnarray}
From our numerical investigations we  obtained the following  statistics for random mixed two-qubit states:
\begin{itemize}
\item Only a tiny fraction $(3.31\pm 0.03)\times 10^{-4}$ violates at least one of the four CHSH inequalities 
that can be constructed from the four Bell-CHSH operators (\ref{AllBell}).
\item Only a small fraction $(1.244 \pm 0.003)\%$ violates at least one of the four RUS inequalities that can be constructed from the same operators
(\ref{AllBell}).
\end{itemize}

We also checked the case of a tomographically complete measurement, where one measures each spin in three fixed spatial directions in each location: from these data 36 Bell-CHSH operators can be constructed. For this sort of measurements we find 
\begin{itemize}
\item The probability to violate at least one of the 36 CHSH inequalities is still small, $(0.249\pm 0.0008)\%$.
\item 
The probability to violate at least one RUS bound out of 36 is $(5.690\pm 0.004)\%$.
\end{itemize}
Finally, what if we construct a set of pure states (randomly distributed according to the Haar measure)? First of all, all pure states, except for a set of measure zero, are entangled. Furthermore, we find
\begin{itemize}
\item
 A fraction $(9.908\pm 0.005)\%$ violates at least one of the four CHSH inequalities that can be constructed from the operators (\ref{AllBell}).
\item A fraction $(46.627 \pm 0.010)\%$ violates  at least one of the four RUS inequalities that can be constructed from the same operators.
\end{itemize}

These statistics (although the exact numbers are measure-dependent, of course) illustrate the fact that violating a CHSH inequality or violating the RUS bound for a fixed correlation measurement is far from necessary  for entanglement (it is always sufficient, though);  being entangled is much less rare than violating a given CHSH or  RUS inequality.

\section*{Acknowledgements}
 This research is supported by the Disruptive Technologies Office (DTO) of the DNI.

\end{document}